\documentclass[12pt,leqno]{article}
\usepackage{graphicx,psfrag,amssymb,amsmath,amsthm}

\newtheorem{theorem}{Theorem}
\newtheorem{lemma}[theorem]{Lemma}

\newtheorem{proposition}[theorem]{Proposition}
\newtheorem{corollary}[theorem]{Corollary}

\def\beq{\begin{equation}}
\def\eeq{\end{equation}}

\def\cB{\mathcal{B}}
\def\cC{\mathcal{C}}

\def\cF{\mathcal{F}}

\def\cK{\mathcal{K}}

\def\cM{\mathcal{M}}

\def\cO{\mathcal{O}}

\def\cT{\mathcal{T}}

\def\IR{{\mathbb R}}

\def\pQ{\partial Q}

\numberwithin{equation}{section}

\begin{document}

\title{A family of chaotic billiards with variable mixing rates}

\author{ N. Chernov$^1$ and H.-K. Zhang$^1$}

\date{\today}

\maketitle

\footnotetext[1]{Department of Mathematics, University of Alabama
at Birmingham;\\ Email:$\ $ chernov@math.uab.edu;
zhang@math.uab.edu}

\begin{abstract}
We describe a one-parameter family of dispersing (hence
hyperbolic, ergodic and mixing) billiards where the correlation
function of the collision map decays as $1/n^a$ (here $n$ denotes
the discrete time), in which the degree $a \in (1, \infty)$
changes continuously with the parameter of the family, $\beta$. We
also derive an explicit relation between the degree $a$ and the
family parameter $\beta$.
\end{abstract}

\centerline{AMS classification numbers: 37D50, 37A25}

\centerline{Keywords: Decay of correlations, dispersing billiards}

\section{Introduction}
\label{secI}

A billiard is a mechanical system in which a point particle moves
in a compact container $Q$ and bounces off its boundary $\pQ$; in
this paper we only consider planar billiards, where $Q \subset
\IR^2$. The billiard dynamics preserves a uniform measure on its
phase space, and the corresponding collision map (generated by the
collisions of the particle with $\pQ$, see below) preserves a
natural (and often unique) absolutely continuous measure on the
collision space. The dynamical properties of a billiard are
determined by the shape of the boundary $\pQ$, and it may vary
greatly from completely regular (integrable) to strongly chaotic.

The dynamics in simple containers (circles, ellipses, rectangles)
are completely integrable. The first class of chaotic billiards
was introduced by Ya.~Sinai in 1970 \cite{Si70}; he proved that if
$\pQ$ is strictly convex inward, its curvature nowhere vanishes,
and the smooth components of $\pQ$ intersect each other
transversally (make no cusps), then the dynamics is hyperbolic
(moreover, uniformly hyperbolic), ergodic, mixing and K-mixing. He
called such systems \emph{dispersing billiards}, now they are
often called \emph{Sinai billiards}. Gallavotti and Ornstein
\cite{GO} proved that Sinai billiards are Bernoulli systems. Later
on the hyperbolicity, ergodicity (as well as Bernoulli property
\cite{CH,OW}) were established for dispersing billiards with cusps
on the boundary \cite{Re} and for billiards whose boundary is
convex (but not strictly convex) inward -- the so called
semi-dispersing billiards -- under certain conditions
\cite{BSC91,CT}.

The rates of mixing (precisely defined in the next section) for
the collision map in dispersing and semidispersing billiards
depend on the shape of the boundary. Assume that\medskip

{\bf (A)} there are no cusps on the boundary and\medskip

{\bf (B)} the boundary curvature does not vanish.\medskip

\noindent Then the collision map is uniformly hyperbolic, and its
mixing properties are very strong -- correlations (defined in the
next section) decay exponentially \cite{Y98,C99}. Relaxing the
requirements (A) and (B) results in nonuniform hyperbolicity and
weaker mixing properties (slower decay of correlations), see
below.

If we relax (A), but not (B), then the correlations appear to
decay polynomially as $\cO(1/n)$. This conjecture is based on
heuristic arguments and numerical experiments \cite{Ma}, and the
work on proving it rigorously is currently underway.

Here we relax (B) but not (A), i.e.\ consider dispersing billiards
without cusps, but assume that the boundary curvature vanishes at
finitely many points (we call them \emph{flat points}). This is a
special class of chaotic billiards hardly ever investigated
before.

First of all, it is easy to show that if there is no periodic
trajectory that hits the boundary at flat points only, then a
certain power of the collision map is uniformly hyperbolic, hence
correlations decay exponentially. In order to weaken the
hyperbolicity and mixing properties, one needs a periodic
trajectory making collisions at flat points only. Then the
vicinity of that periodic orbit acts as a ``trap'' where
hyperbolicity may remain weak for arbitrarily long times.

For simplicity we assume that there is one such periodic
trajectory of period two that runs between two flat points. More
precisely, let the boundary $\pQ$ near those two flat points be
given by the equations
\beq \label{xy}
         y = \pm g_{\beta}(x),\qquad
         g_{\beta}(x)=|x|^\beta + 1 \qquad (\beta>2)
\eeq
in some rectangular coordinate system in $\IR^2$. The billiard
table lies between the ``$+$'' and ``$-$'' branches of the above
function, and elsewhere it is bounded by ``regular dispersing''
curves, which are strictly convex inward with nowhere vanishing
curvature and make no cusps. Note that the curvature of the
boundary does vanish at the points $(0,1)$ and $(0,-1)$, because
$\beta >2$, and the periodic orbit runs between these points along
the $y$ axis. The power $\beta >2$ is the parameter of the so
constructed family of billiard tables.

Our main result, stated precisely in the next section, is that the
correlations for the collision map decay as $\cO(1/n^a)$, where
$$
        a = \frac{\beta+2}{\beta-2}.
$$
Therefore, the degree $a$ covers the entire interval from one to
infinity. In the limit $\beta \to \infty$, the boundary flattens
out, and the correlations decay almost as $1/n$, which is an
established result for semi-dispersing billiards with two parallel
flat components of the boundary \cite{CZ}. In the limit $\beta \to
2$, the boundary ``curves up'' and approaches strictly dispersing
case $y=\pm(x^2+1)$ with nowhere vanishing curvature; then $a \to
\infty$ and so the correlations decay faster than any polynomial
function. In the limit $\beta=2$ the correlations decay
exponentially \cite{C99}. Thus by varying the parameter $\beta$ we
can adjust the degree of the polynomial decay rate $1/n^a$ to any
value $a \in (1,\infty)$.

\section{Statement of results}
\label{secSR}

First we recall standard definitions of billiard theory
\cite{BSC90,BSC91,C97,C99}. A billiard is a dynamical system where
a point moves freely at unit speed in a domain $Q$ ({\em the
table}) and reflects off its boundary $\pQ$ ({\em the wall}) by
the rule ``the angle of incidence equals the angle of
reflection''. We assume that $Q\subset\IR^2$ and $\pQ$ is a finite
union of $C^3$ curves (arcs). The phase space of this system is a
three dimensional manifold $Q\times S^1$. The dynamics preserves a
uniform measure on $Q\times S^1$.

\begin{figure}[h]
\centering \centering \psfrag{r}{$r$} \psfrag{p}{$\varphi$}
\psfrag{0}[][]{$\varphi=0$} \psfrag{1}{$\frac{\pi}{2}$}
\psfrag{2}{$-\frac{\pi}{2}$} \psfrag{a}{$(a)$} \psfrag{b}{$(b)$}
\includegraphics[height=1.5in]{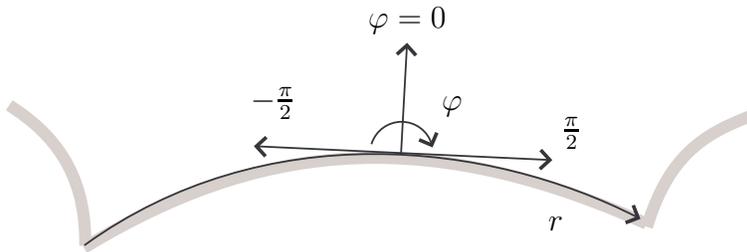}
\renewcommand{\figurename}{Fig.}
\caption{Orientation of $r$ and $\varphi$}\label{rp}
\end{figure}

Let $\cM=\pQ\times [-\pi/2,\pi/2]$ be the standard cross-section
of the billiard dynamics, we call $\cM$ the {\em collision space}.
Canonical coordinates on $\cM$ are $r$ and $\varphi$, where $r$ is
the arc length parameter on $\pQ$ and $\varphi\in [-\pi/2,\pi/2]$
is the angle of reflection, see Fig.~\ref{rp}. We denote by $\pi$
the natural projection of $M$ onto $\pQ$.

The first return map $\cF:\cM\to \cM$ is called the {\em collision
map} or the {\em billiard map}, it preserves smooth measure
$d\mu=\cos\varphi\, dr\, d\varphi$ on $\cM$.

Let $f,g\in L^2_{\mu}(\cM)$ be two functions. {\em Correlations}
are defined by
\beq
   \cC_n(f,g,\cF,\mu) = \int_{\cM} (f\circ \cF^n)\, g\, d\mu -
    \int_{\cM} f\, d\mu    \int_{\cM} g\, d\mu
       \label{Cn}
\eeq
It is well known that $\cF:\cM\to\cM$ is {\em mixing} if and only
if
\beq \label{Cto0}
   \lim_{n\to\infty}
   \cC_n(f,g,\cF,\mu) = 0
   \qquad \forall  f,g\in L^2_{\mu}(\cM)
\eeq
The rate of mixing of $\cF$ is characterized by the speed of
convergence in (\ref{Cto0}) for smooth enough functions $f$ and
$g$. We will always assume that $f$ and $g$ are H\"older
continuous or piecewise H\"older continuous with singularities
that coincide with those of the map $\cF^k$ for some $k$. For
example, the length of the free path between successive
reflections is one such function.

We say that correlations decay {\em exponentially} if
$$
    |\cC_n(f,g,\cF,\mu)|<\,{\rm const}\cdot e^{-cn}
$$
for some $c>0$ and {\em polynomially} if
$$
   |\cC_n(f,g,\cF,\mu)|<\,{\rm const}\cdot n^{-a}
$$
for some $a>0$. Here the constant factor depends on $f$ and $g$.

Next we state our results.

\begin{figure}[h]
\centering \centering \psfrag{Q}{$Q$}
\psfrag{a}[][]{$y=|x|^\beta+1$} \psfrag{b}[][]{$y=-(|x|^\beta+1)$}
\psfrag{x}{\small$x$} \psfrag{y}{\small $y$}
\includegraphics[height=1.6in]{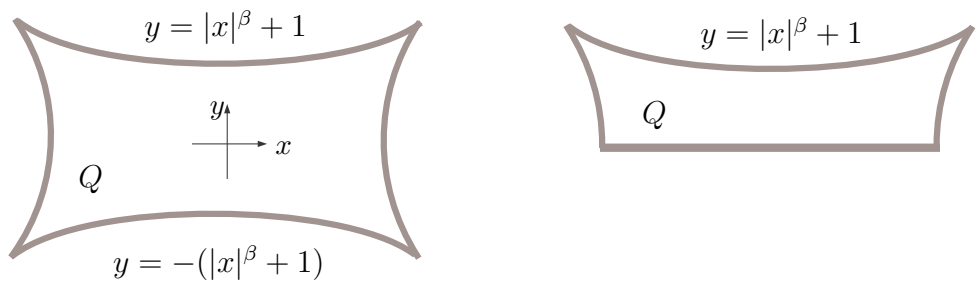}
\renewcommand{\figurename}{Fig.}
\caption{Dispersing billiards with walls where the curvature
vanishes.}\label{curvature1}
\end{figure}

Let $Q \subset \IR^2$ be a domain bounded by the curves
$y=g_{\beta}(x)$ and $y=-g_{\beta}(x)$, see (\ref{xy}), and
several strictly convex (inward) curves with nowhere vanishing
curvature and no cusps. An example is shown on
Fig.~\ref{curvature1} (left).

Our results also apply to billiards bounded by one of the curves
(\ref{xy}), say $y=g_{\beta}(x)$, the $x$-axis and several
strictly convex (inward) curves with nowhere vanishing curvature
and no cusps. An example is shown on Fig.~\ref{curvature1}
(right).

\begin{theorem} \label{TmMain}
For the above  billiard tables, the correlations
(\ref{Cn}) for the billiard map $\cF:\cM\to\cM$ and piecewise
H\"older continuous functions $f,g$ on $\cM$ decay as
\beq \label{main}
   |\cC_n(f,g,\cF,\mu)|\leq\,{\rm const}\cdot
   \frac{(\ln n)^{a+1}}{n^{a}},
\eeq
where $a = (\beta+2)/(\beta-2)$.
\end{theorem}

\noindent{\em Remark}. The logarithmic factor in (\ref{main}) is a
by-product of a general method for correlation analysis developed
in \cite{M04,CZ}. Perhaps it is possible to suppress it by using
more powerful Young's techniques \cite{Y99} but this may require a
substantial extra effort.

\section{Proof of the main Theorem}
\label{secPMT}

We use a general scheme for the analysis of hyperbolic dynamical
systems with polynomial decay of correlations developed in
\cite{CZ} (which is an extension of earlier works by Young
\cite{Y99} and Markarian \cite{M04}). That scheme has been
successfully applied in \cite{CZ} to various classes of chaotic
billiards.

The scheme is based on finding a subset $M \subset \cM$ where the
map $\cF$ is strongly (uniformly) hyperbolic and the subsequent
analysis of the return map $F\colon M \to M$, which is defined by
\beq \label{Fdef}
   F(X) = \cF^{N(X)}(X), \qquad
   N(X) = \min\{i>0 \colon \cF^i(X)\in M\}.
\eeq

\begin{figure}[h]
\centering \centering \psfrag{e}{$\varepsilon$} \psfrag{1}{$q_1$}
\psfrag{2}{$q_2$} \psfrag{3}{$q_3$}
\psfrag{4}{$q_4$}\psfrag{y}{$y$}
\includegraphics{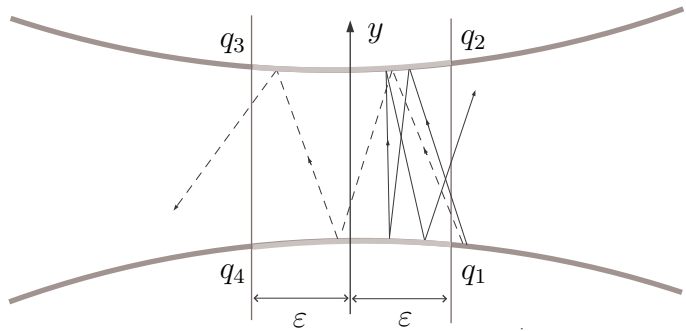}
\renewcommand{\figurename}{Fig.}
\caption{A window cut through $\pQ$ to construct
$M$.}\label{FigWindow}
\end{figure}

In our case the hyperbolicity is strong everywhere except the
vicinity of the two flat points $(0,1)$ and $(0,-1)$ on $\pQ$. We
fix an $\varepsilon >0$ and define
$$
   M = \bigl(\pQ \setminus \{|x|<\varepsilon\}\bigr)
   \times [-\pi/2,\pi/2] =
   \cM \setminus \pi^{-1}(\{|x|<\varepsilon\}),
$$
i.e. we remove from $\pQ$ a narrow window -- the
$\varepsilon$-neighborhood of the $y$ axis -- that contains both
flat points, see Fig.~\ref{FigWindow}. Let $q_1 = (\varepsilon,
-g_{\beta}(\varepsilon))$ denote one of the four points on $\pQ$
that border the window $|x| < \varepsilon$, and by $q_2, q_3, q_4$
the other three points (Fig.~\ref{FigWindow}).

In order to prove Theorem~\ref{TmMain}, according to our general
scheme \cite{CZ}, we need to establish two properties of the map
$F$ described below. \medskip

\noindent {\bf (F1)} First, the map $F\colon M\to M$ enjoys
exponential decay of correlations. Moreover, there is a horseshoe
$\Lambda \subset M$ with a hyperbolic structure such that the
return times to $\Lambda$ obey an exponential tail bound, see
\cite{Y98,Y99,CZ} for precise definitions. \medskip

\noindent {\bf (F2)} Second, the return times to $M$ under the
original map $\cF$ defined by
$$
     R(X;\cF,M)=\min\{r\geq 1:\ \cF^r(X)\in M\}
$$
satisfy the polynomial tail bound
\beq \label{Mpol}
    \mu(X\in M:\ R(X;\cF,M)>n)\leq
    \, {\rm const}\cdot n^{-a-1}\quad\quad \forall n\geq 1
\eeq
where $a>0$ is the constant of Theorem~\ref{TmMain}. \medskip

\noindent It is shown in \cite{CZ} that Theorem~\ref{TmMain}
follows from (F1) and (F2). Also, the proof of (F1) is reduced in
\cite{CZ} to the verification of the following property of
unstable manifolds:

Let $W \subset M$ denote an unstable manifold (it is a smooth
curve since dim$\, M =2$). Since the map $F$ has singularities
(described below) the image $F(W)$ may consist of finitely or
countably many unstable manifolds. Let $W_i$, $i\geq 1$, denote
the preimages of the smooth components of $F(W)$, i.e.\ the
subcurves $W_i \subset W$ on which the map $F$ is smooth. Next,
for every point $X\in W_i$ denote by $\Lambda(X)$ the Jacobian of
the map $F$ restricted to $W_i$, i.e.\ the local factor of
expansion (stretching factor) of the curve $W_i$ under the map $F$
at the point $X$. Put
$$
   \Lambda_i = \min_{X\in W_i} \Lambda(X)
$$
In order to prove (F1) we need to verify that
\beq
   \liminf_{\delta\to 0}\
   \sup_{W\colon |W|<\delta} \sum_i \Lambda_i^{-1} <1,
      \label{step1}
\eeq
where the supremum is taken over unstable manifolds $W$ of length
$<\delta$.

The reduction of (F1) to (\ref{step1}) is carried out in \cite{CZ}
for very general 2D hyperbolic maps that include our family of
dispersing billiards.

Thus it remains to prove (F2) and (\ref{step1}). This requires
detailed investigation of the singularities of the map $F$. The
definition (\ref{Fdef}) makes it clear that $F$ is singular at $X$
whenever $\cF(X)$ or $N(X)$ is singular. The singularities of the
original map $\cF$ are well studied \cite{C99} and the estimate
(\ref{step1}) is proved for unstable manifolds affected by those,
so we focus on the singularities of $N(X)$.

The value $n = N(X)-1$ is the number of bounces the billiard
trajectory of the point $X\in M$ experiences in the window $|x| <
\varepsilon$ before returning to $M$. For large $N(X)$, the
trajectory of $X$ runs almost parallel to the $y$ axis for a long
time, and we distinguish two types of such trajectories, see
Fig.~\ref{FigWindow}. The trajectories of the first type enter the
window, almost approach its central axis (the $y$ axis), but then
turn back and exit on the same side they entered (the solid line
on Fig.~\ref{FigWindow}). The trajectories of the other type move
through the window, cross the $y$ axis, and exit on the opposite
side (the dashed line on Fig.~\ref{FigWindow}). These two types of
trajectories are separated by points whose trajectories converge
to the $y$ axis and never return to $M$. The singularities of
$N(X)$ occur at points where the number of bounces in the window
$|x| < \varepsilon$ changes from $n$ to $n+1$ or $n-1$.

\begin{figure}[h]
\centering \psfrag{M}{$M$} \psfrag{1}{$S_1$} \psfrag{n}{$S_n''$}
\psfrag{m}{$S_n'$} \psfrag{o}{$1/n^b$} \psfrag{E}{$E$}
\psfrag{S}[][]{$S_{\infty}$}
\includegraphics[height=3in]{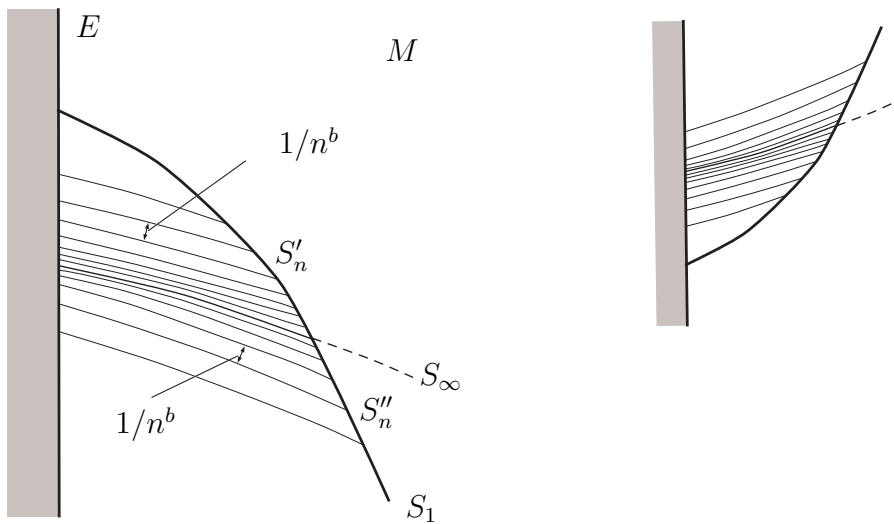}
\renewcommand{\figurename}{Fig.}
\caption{Singularities of the map $F$ (left) and $F^{-1}$
(right).}\label{FigSing}
\end{figure}

Figure~\ref{FigSing} (left) shows the structure of singularity
lines of the map $F$ near the point $q_1$, in the $r,\varphi$
coordinates. The bold vertical line $E$ on the left is
$\pi^{-1}(q_1)$, the edge of $M$. The bold steeply decreasing
curve $S_1$ terminating on $E$ consists of points $\{X\colon
\cF(X)\in\pi^{-1}(q_2)\}$, which hit the point $q_2$ of $\pQ$
under the map $\cF$. The points above $S_1$ are mapped by $\cF$ to
the right of $q_2$, so they do not leave $M$. The points below the
curve $S_1$ are mapped by $\cF$ to the left of $q_2$, and then
they enter the window $|x| < \varepsilon$.

The decreasing curve $S_{\infty}$ which crosses $S_1$ and
terminates on $E$ consists of points whose trajectories converge
to the $y$ axis (thus $S_{\infty}$ is the stable manifold of the
periodic orbit running along the $y$ axis). The dashed part of
$S_{\infty}$ (to the right of $S_1$) does not enter the window
immediately, but will do so in one or a few iterations.

The region above $S_{\infty}$ but below $S_1$ consists of points
whose trajectories enter the window but turn back without reaching
the $y$ axis (like the solid trajectory on Fig.~\ref{FigWindow}).
This region is divided into infinitely many strips by decreasing
curves $S_n'$, $n\geq 1$, which correspond to the discontinuities
of the function $N(X)$: the curve $S_n'$ separates the region
$C_n'\colon = \{N(X)=n\}$ from the similar region $C_{n+1}'$. The
curves $S_n'$ are almost parallel to $S_{\infty}$ and accumulate
toward $S_{\infty}$ from above.

The region below $S_{\infty}$ consists of points whose
trajectories enter the window and manage to move through it
crossing the $y$ axis (like the dashed trajectory on
Fig.~\ref{FigWindow}). This region is divided into infinitely many
strips by decreasing curves $S_n''$, $n\geq 1$, which correspond
to the discontinuities of the function $N(X)$: the curve $S_n''$
separates the region $C_n''\colon = \{N(X)=n\}$ from the similar
region $C_{n+1}''$. The curves $S_n''$ are almost parallel to
$S_{\infty}$ and accumulate toward $S_{\infty}$ from below.

Due to the time-reversibility of the billiard dynamics, the
singularities of the map $F^{-1}$ have a similar structure. In
fact, the picture shown on Fig.~\ref{FigSing} (left) must be
flipped about the horizontal line $\varphi=0$ to become the
illustration of singularity curves of $F^{-1}$ near the same point
$q_1$, see a scaled-down version of it shown on Fig.~\ref{FigSing}
(right).

\begin{figure}[h]
\centering \psfrag{F}{$F$} \psfrag{C}{$C_n'$} \psfrag{W}{$W$}
\psfrag{I}{$F(W)$}
\includegraphics[width=5in]{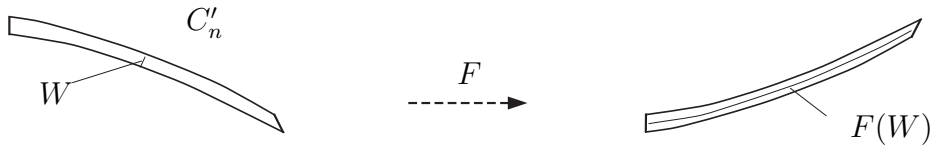}
\renewcommand{\figurename}{Fig.}
\caption{The transformation of $C_n'$ under $F$.}\label{FigCF}
\end{figure}

Furthermore, $F$ maps each region $C_n'$ onto a symmetric region
made by the singularity curves of $F^{-1}$ (near the point $q_1$
or $q_2$). Similarly, $F$ maps each region $C_n''$ onto a
symmetric region made by the singularity curves of $F^{-1}$ near
the point $q_3$ or $q_4$. The action of $F$ on $C_n'$ is
schematically shown on Fig.~\ref{FigCF}: long sides of $C_n'$ are
transformed into short sides of $F(C_n')$, while short sides of
$C_n'$ are transformed into long sides of $F(C_n')$. Unstable
manifolds $W\subset C_n'$ (which are short increasing curves in
the $r,\varphi$ coordinates) are mapped onto long unstable curves
stretching across $F(C_n')$ completely, see Fig.~\ref{FigCF}. Let
$h_n$ denote the height of the region $C_n$ (of course, it is not
uniform across $C_n'$, but we can take the maximum height, for
example). Then, since the length of $F(C_n')$ is $\cO(1)$, the
factor of expansion of unstable manifolds $W\subset C_n'$
is\footnote{Our notation $A\sim B$ has the following meaning:
there is a constant $C = C(Q)>1$ such that $C^{-1} < A/B < C$.}
\beq \label{Lambdah}
      \Lambda_n \sim 1/h_n.
\eeq
(this, of course, requires the distortions be uniformly bounded on
$W \subset C_n'$, which follows from general results
\cite{C99,CZ}). A similar analysis applies to the region $C_n''$.

The qualitative description of the singularity curves for the map
$F$ outlined above is the result of rather straightforward (albeit
somewhat meticulous) geometric considerations, which we omit. In
order to determine the rates of the decay of correlations we need
certain quantitative estimates on the measure of the regions
$C_n'$ and $C_n''$ and on the factor of expansion of unstable
manifolds $W \subset C_n'$ and $W\subset C_n''$ under the map $F$.

\begin{proposition} \label{PrAux}
Unstable manifolds $W\subset C_n'$ and $W\subset C_n''$ are
expanded under the map $F$ by a factor $\Lambda_n \sim n^b$, where
$b=a+2$. Accordingly, see (\ref{Lambdah}), the height (and hence
the measure) of the regions $C_n'$ and $C_n''$ is $\sim n^{-b}$.
\end{proposition}

The proposition will be proven in the next section. Here we
complete the proof (F2) and (\ref{step1}), thus deriving
Theorem~\ref{TmMain}.

It is immediate that
$$
    \mu(X\in M:\ R(X;\cF,M)>n)
    =\sum_{m> n} \mu(C_m'\cup C_m'')\leq
    \, {\rm const}\cdot n^{-a-1}
$$
which proves (F2).

Next, every unstable manifold $W \subset M$ is a smooth
monotonically increasing curve in the $r,\varphi$ coordinates.
Hence for every $n\geq 1$ the intersection $W\cap C_n'$ is at most
one curve, and the same is true for $W\cap C_n'$. If $W$ crosses
the separating line $S_{\infty}$, then it intersects $C_n'$ and
$C_n''$ for all $n\geq n_{\delta}$, where $n_{\delta}$ grows to
$\infty$ as $|W|=\delta$ converges to $0$. Then
$$
   \sum_i \Lambda_i^{-1} <
   \,{\rm const}\, \sum_{n=n_{\delta}}^{\infty} \frac{1}{n^{a+2}}
   <\frac{\rm const}{n_{\delta}^{a+1}},
$$
which is less than $1$ for all sufficiently small $\delta>0$. If
$W$ does not cross $S_{\infty}$, but crosses $S_n'$ or $S_n''$
with sufficiently large $n$, the analysis is similar. If $W$ only
crosses $S_n'$ or $S_n''$ with small $n$, then a standard trick -
the use of a higher iterate of $F$ -- applies, see \cite{CZ}. \qed
\medskip

\noindent{\em Remark}. To establish an upper bound on
correlations, we only need an upper bound on the measures in
(\ref{Mpol}). Thus it will be enough to obtain a lower bound on
$\Lambda_n$ in Proposition~\ref{PrAux}. This is what we do in the
next section: we prove that $\Lambda_n \geq \,\text{const}\, n^b$.
While our arguments can be easily extended to obtain an upper
bound $\Lambda_n \leq \,\text{const}\, n^b$ as well, we do not
pursue this goal.

\section{Proof of Proposition~\ref{PrAux}}
\label{secPP}

Given an unstable manifold $W \subset C_n'$ (or $W \subset C_n''$)
and a point $X \in W$, the map $F = \cF^n$ expands $W$ at $X$ by
the factor \cite{BSC91,C99}
\beq \label{Lambdan}
  \Lambda_n(X) = \prod_{m=0}^{n-1} \bigl(1 + \tau(X_m) \cB(X_m)\bigr)
\eeq
where $X_m = \cF^m (X)$, and for every point $Y =(r,\varphi) \in
\cM$ we denote by $\tau(Y)$ the time between the collisions at the
points $Y$ and $\cF(Y)$ and
\beq \label{cBY}
    \cB(Y) = \frac{1}{\cos\varphi}\,
    \biggl(\frac{d\varphi}{dr}+\cK(r)\biggr),
\eeq
where $d\varphi/dr$ denotes the slope of the unstable manifold
$W(Y)$ passing through $Y$ and $\cK (r)$ the curvature of the
boundary $\pQ$ at the point $r$.

We note that $\cB(Y)$ is the geometric curvature of the orthogonal
cross-section of the family of trajectories on the billiard table
$Q$ coming from $W(Y)$, see \cite{BSC90,BSC91,C99} for more
details. The expansion factor (\ref{Lambdan}) is measured in the
so called p-norm defined by
\beq \label{pnorm}
   |V|_p = \cos\varphi\, |dr|
\eeq
for tangent vectors $V=(dr,d\varphi) \in \cT_X\cM$. The p-norm is
equivalent to the Euclidean norm
\beq \label{Enorm}
    |V|=\bigl[(dr)^2+(d\varphi)^2\bigr]^{1/2}
\eeq
along the trajectory of $\cF^m(X)$, $1\leq m\leq n$, as we will
prove below.

The value of $\cB (Y)$ is positive for all $Y \in \cM$. The
initial value $\cB (X)$, $X \in W$, is bounded away from zero and
infinity:
$$
    \cB_{\min} \leq \cB(X) \leq \cB_{\max},
$$
where $\cB_{\min} >0$ is determined by our choice of
$\varepsilon$. For the computation of $\cB (X_m)$ we have a
recurrent formula
\beq \label{cBXm}
   \cB(X_{m}) = \frac{2\cK(r_{m})}{\cos\varphi_{m}}
   +\frac{1}{\tau (X_{m-1}) + 1/\cB(X_{m-1})},
\eeq
where $(r_m, \varphi_m) = X_m$. Let $x_m$ denote the $x$
coordinate of the collision point $r_m \in \pQ$, then it is easy
to compute
\beq \label{cKrm}
    \cK(r_m) = \frac{\beta(\beta-1)|x_m|^{\beta-2}}
    {\bigl(1+\beta^2|x_m|^{2(\beta-1)}\bigr)^{3/2}}.
\eeq
We note that $\cK (r_m)$ approaches zero, as $x_m$ approaches
zero, and we will see later that $\cB (X_m)$ approach zero as
well.

\begin{figure}[h]
\centering \psfrag{q}[][]{$x_{m+1}$} \psfrag{p}[][]{$x_{m}$}
\psfrag{x}{$x$} \psfrag{y}{$y$} \psfrag{Q}{$Q$}
\psfrag{vm}{$w_{m+1}$} \psfrag{vm-1}{$w_{m}$}
\includegraphics[height=2in]{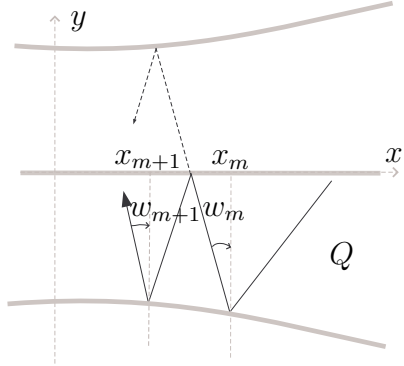}
\renewcommand{\figurename}{Fig.}
\caption{The m-th collision and the parameters}\label{Figxw}
\end{figure}

Next we consider the trajectory of a point $X\in C_n''$ (the case
$X \in C_n'$ is easier and will be treated later). Due to an
obvious symmetry of the table $Q$ about the $x$-axis it is
convenient to fold $Q$ in half and reflect its upper part $y>0$
onto its lower half $y<0$, then our trajectory will bounce between
the $x$-axis and the lower side of $Q$, see Fig.~\ref{Figxw}.

Let $n'$ be uniquely defined by $x_{n'+1}<0<x_{n'}$. First we
consider the interval $1\leq m\leq n'$, i.e.\ where $x_m>0$.

We denote by $w_m$ the angle made by the $y$-axis and the velocity
vector after the $m$th collision. Note that $(\beta x^{\beta-1},
1)$ is the inward normal vector to $\pQ$ at the point $r_m$.
Elementary geometric considerations yield the following relations:
\beq \label{wxm}
  \begin{split}
  w_{m}-w_{m+1} &= 2\arctan (\beta x^{\beta-1}_{m+1})\\
  x_{m}-x_{m+1} &= 2\tan w_m+(x_m^{\beta}+x_{m+1}^{\beta})\tan
  w_m.
  \end{split}
\eeq
Using Taylor expansion we obtain
\beq \label{wxm1}
  \begin{split}
    w_{m}-w_{m+1} &= 2\beta x_{m+1}^{\beta-1} - R_{w, m+1} \\
    x_{m}-x_{m+1} &= 2w_m + R_{x, m+1},
  \end{split}
\eeq
where
\beq \label{R1}
  R_{w, m+1}=\tfrac 23 \beta^3 x_{m+1}^{3(\beta-1)}
  +\cO(x_{m+1}^{5(\beta-1)}) > 0
\eeq
and
\beq \label{R2}
   R_{x, m+1}=\tfrac 23 w_m^3 + (x_m^{\beta}+x_{m+1}^{\beta})w_m
  +\cO(x_m^{\beta}w_m^3+w_m^5) > 0
\eeq
(the positivity of $R_{w, m+1}$ and $R_{x, m+1}$ is guaranteed by
the smallness of $\varepsilon$). Note that both $\{x_m\}$ and
$\{w_m\}$ are decreasing sequences of positive numbers for
$m=1,\ldots,n'$.

\begin{lemma}
Let $n''\in [1,n']$ be uniquely defined by the condition
\beq \label{n''}
     w_{n''-1}>2w_{n'}>w_{n''}.
\eeq
Then for all $n'' \leq m \leq n'$
\beq \label{xmasym}
   x_m \sim (n'-m)w_{n'}
\eeq
and
\beq \label{n'n''asym}
   n'-n'' \sim w_{n'}^{\frac{2-\beta}{\beta}}
\eeq
(recall our convention on the usage of ``$\sim$'' in the previous
section).
\end{lemma}

\proof Due to (\ref{wxm1}) and (\ref{n''}), for any $m\in
[n'',n')$ we have
$$
   2w_{n'}\leq 2w_m \leq x_m-x_{m+1} \leq 3w_m \leq 6w_{n'}
$$
hence
\beq \label{xmn''}
   2(n'-m)w_{n'} \leq x_m \leq 6(n'-m+1)w_{n'}
\eeq
(note that $0\leq x_{n'} \leq 3w_{n'}$). Next, due to
(\ref{xmn''}) and (\ref{wxm1})
$$
     2^{\beta-1}(n'-m-1)^{\beta-1}w_{n'}^{\beta-1}\leq
     w_m-w_{m+1}\leq
     2\beta\, 6^{\beta-1}(n'-m)^{\beta-1}w_{n'}^{\beta-1}
$$
therefore
$$
       w_m \sim w_{n'}+(n'-m)^{\beta} w_{n'}^{\beta-1}
$$
Substituting $m=n''$, then $m=n''-1$ and using (\ref{n''}) implies
(\ref{n'n''asym}) and completes the proof of the lemma. \qed
\medskip

We note that (\ref{xmasym}) and (\ref{n'n''asym}) imply
\beq \label{wxn''}
    x_{n''} \sim w_{n'}^{\frac{2}{\beta}},
\eeq
hence $x_{n''} \ll \varepsilon$ and thus $n'' \gg 1$. Next we
consider the case $1< m\leq n''$.

\begin{lemma}
For all $1< m \leq n''$ we have
\beq \label{xwbeta}
   x_{m}^{\beta} \sim w_{m}^{2} \sim m^{\frac{2\beta}{2-\beta}}.
\eeq
Moreover,
\beq \label{n''n'}
   n'' \sim w_{n'}^{\frac{2-\beta}{\beta}}.
\eeq
\end{lemma}

\proof Due to (\ref{wxm1}) and the mean value theorem, for some
$x_{\ast}\in (x_{m+1},x_m)$
\begin{align*}
   x_m^{\beta}-x_{m+1}^{\beta} &=
   \beta x_{\ast}^{\beta-1}(x_m-x_{m+1}) \\
   &=2\beta x_{\ast}^{\beta-1}w_m +
   \cO\bigl(x_m^{\beta-1}w_m^3+x_m^{2\beta-1}w_m\bigr).
\end{align*}
Similarly, for some $w_{\ast}\in (w_{m+1},w_m)$
\begin{align*}
   w_m^{2}-w_{m+1}^{2} &=
   2w_{\ast}(w_m-w_{m+1}) \\
   &=4\beta x_{m+1}^{\beta-1}w_{\ast} +
   \cO\bigl(x_m^{3(\beta-1)}w_m\bigr).
\end{align*}
This easily implies
$$
   1\leq \frac{w_m^{2}-w_{m+1}^{2}}
   {x_m^{\beta}-x_{m+1}^{\beta}} \leq 5.
$$
Also, (\ref{wxn''}) and (\ref{n''}) imply that $w_{n''}^2 \sim
x_{n''}^{\beta}$, i.e.
$$
     C'\leq \frac{w_{n''}^{2}}
     {x_{n''}^{\beta}} \leq C''
$$
where we can assume $C'<1$ and $C''>5$. Now the first relation in
(\ref{xwbeta}) follows easily.

Next, denote $z_m = x_m^{\frac{\beta-2}{2}}$. Then (\ref{wxm1})
and the mean value theorem imply
\begin{align*}
   z_m - z_{m+1}
   &\sim x_m^{\frac{\beta-4}{2}}(x_m-x_{m+1})\\
   &\sim x_m^{\frac{\beta-4}{2}}w_m \\
   &\sim x_m^{\beta-2} =z_m^2
\end{align*}
(we used the first relation in (\ref{xwbeta})). Now let $Z_m =
1/z_m$, then
$$
   Z_{m+1} - Z_m \sim Z_{m+1}/Z_m \sim 1
$$
(we note that $x_{m}-x_{m+1} \sim x_m^{\frac{\beta}{2}} \ll x_m$,
hence $x_m/x_{m+1} \approx 1$). Since $x_0 \geq \varepsilon$,
\beq \label{Z0}
     Z_0 \leq \varepsilon^{-\frac{\beta-2}{2}}=\,\text{const},
\eeq
and we obtain
\beq \label{Zm}
   Z_m \sim m
   \qquad\text{and}\qquad
   z_m \sim 1/m,
\eeq
which proves the second relation in (\ref{xwbeta}). Now
(\ref{n''n'}) is immediate due to (\ref{wxn''}). \qed

Equations (\ref{n'n''asym}) and (\ref{n''n'}) imply $n'' \sim
n'-n''$ and $n' \sim w_{n'}^{\frac{2-\beta}{\beta}}$. A similar
analysis can be done for the remaining part of the trajectory, $n'
< m <n$, which shows that $n-n' \sim
w_{n'+1}^{\frac{2-\beta}{\beta}}$. Since $w_{n'} \approx
w_{n'+1}$, we obtain $n-n' \sim n'$, and so
\beq \label{nnn}
   n'' \sim n
   \qquad\text{and}\qquad
   w_{n'} \sim n^{\frac{\beta}{2-\beta}}.
\eeq

\begin{lemma} \label{LmMonot}
For all $m < n'$ we have
$$
   w_m^2-2 x_m^\beta <
   w_{m+1}^2-2 x_{m+1}^\beta
$$
i.e.\ $\{w_m^2-2 x_m^\beta\}$ is an increasing sequence for
$m=1,\ldots,n'$.
\end{lemma}

\proof By the convexity of the function $x^{\beta}$,
$$
  2x_{m}^\beta - 2x_{m+1}^\beta \geq 2\beta x_{m+1}^{\beta-1}
  (x_m-x_{m+1}).
$$
Now due to (\ref{wxm1})--(\ref{R2})
$$
  \quad\qquad 2\beta x_{m+1}^{\beta-1}(x_m-x_{m+1})
  > 2w_m (w_m-w_{m+1}) > w_m^2 - w_{m+1}^2. \quad\qquad \qed
$$

Lemma~\ref{LmMonot} implies $w_m^2 -2x^{\beta}_m <w_{n'}^2$, hence
\beq \label{monot}
  w_m < \sqrt{2x_m^\beta + w_{n'}^2}
  < \sqrt{2}\, x_m^{\frac{\beta}{2}} +
   \tfrac 12 \, x_m^{-\frac{\beta}{2}} w_{n'}^2.
\eeq

Next we derive a more precise estimate on the $x$ coordinate:

\begin{lemma}
For all $1\leq m \leq n''$ we have
\beq \label{xmain}
   x_m^{\frac{2-\beta}{2}} \leq
   L m + C_1 \ln m + C_2 m
   \Bigl( \frac mn \Bigr )^{\frac{2\beta}{\beta-2}} + C_3
\eeq
where $L=(\beta-2)\sqrt{2}$ and $C_1,C_2,C_3>0$ are some
constants.
\end{lemma}

\proof Due to (\ref{wxm1}) and (\ref{monot})
$$
  x_m-x_{m+1} <
  2\sqrt{2}\, x_m^{\frac{\beta}{2}}
  + x_m^{-\frac{\beta}{2}} w_{n'}^2
  + Cx_m^{\frac{3\beta}{2}}
$$
for some large $C>0$ (we used the fact $w_m^2 \sim x_m^{\beta}$).
As before, we put $z_m = x_m^{\frac{\beta-2}{2}}$. We consider two
cases. If $\beta \geq 4$, then the function
$x^{\frac{\beta-2}{2}}$ is convex down, and
\begin{align*}
   z_m - z_{m+1}
   &\leq \tfrac{\beta-2}{2}\,x_m^{\frac{\beta-4}{2}}(x_m-x_{m+1})\\
   &\leq L\, x_m^{\beta-2} +
   \tfrac{\beta-2}{2}\,x_m^{-2}w_{n'}^2 + Cx_m^{2\beta-2}\\
   &\leq L\, z_m^{2} +
   \tfrac{\beta-2}{2}\,z_m^{-\frac{4}{\beta-2}}w_{n'}^2
   + Cz_m^{\frac{4\beta-4}{\beta-2}}.
\end{align*}
If $\beta < 4$, then the function $x^{\frac{\beta-2}{2}}$ is
convex up, and
\begin{align*}
   z_m - z_{m+1}
   &\leq \tfrac{\beta-2}{2}\,x_{m+1}^{\frac{\beta-4}{2}}(x_m-x_{m+1})\\
   &\leq L\, x_m^{\frac{\beta}{2}}\, x_{m+1}^{\frac{\beta-4}{2}} +
   \tfrac{\beta-2}{2}\,x_{m+1}^{-2}w_{n'}^2 + Cx_m^{2\beta-2}\\
   &\leq L\, z_m^{\frac{\beta}{\beta-2}}\, z_{m+1}^{\frac{\beta-4}{\beta-2}} +
   \tfrac{\beta-2}{2}\,z_{m+1}^{-\frac{4}{\beta-2}}w_{n'}^2
   + Cz_m^{\frac{4\beta-4}{\beta-2}}
\end{align*}
As before, let $Z_m = 1/z_m$. Then in the case $\beta>4$ we
have
\begin{align*}
   Z_{m+1} - Z_{m}
   &\leq L\, \frac{Z_{m+1}}{Z_m} +
   \frac{\beta-2}{2}\,Z_{m+1}^{\frac{2\beta}{\beta-2}}w_{n'}^2
   + CZ_{m+1}^{-\frac{2\beta}{\beta-2}} \\
   &\leq L + L\, \frac{Z_{m+1}-Z_m}{Z_m} +
   \frac{\beta-2}{2}\,Z_{m+1}^{\frac{2\beta}{\beta-2}}w_{n'}^2
   + CZ_{m+1}^{-\frac{2\beta}{\beta-2}}
\end{align*}
Solving the last inequality for $Z_{m+1} - Z_{m}$ and using
(\ref{Zm}) and (\ref{nnn}) gives
$$
   Z_{m+1} - Z_{m} \leq L+\frac{C'}{m} + C''
   \Bigl( \frac mn \Bigr )^{\frac{2\beta}{\beta-2}}
$$
for some large $C',C''>0$. Summing up over $m$ implies
(\ref{xmain}) with $C_3=\varepsilon^{-\frac{\beta-2}{2}}$, see
(\ref{Z0}).

In the other case, $\beta<4$, we have
\begin{align*}
   Z_{m+1} - Z_{m}
   &\leq L\biggl(\frac{Z_{m+1}}{Z_m}\biggr)^{\frac{2}{\beta-2}}+
   \frac{\beta-2}{2}\,Z_{m+1}^{\frac{2\beta}{\beta-2}}w_{n'}^2
   + CZ_{m+1}^{-\frac{2\beta}{\beta-2}} \\
   &\leq L + G\, \frac{Z_{m+1}-Z_m}{Z_m} +
   \frac{\beta-2}{2}\,Z_{m+1}^{\frac{2\beta}{\beta-2}}w_{n'}^2
   + CZ_{m+1}^{-\frac{2\beta}{\beta-2}}
\end{align*}
with $G=\tfrac{3L}{\beta-2}$, and the subsequent analysis is
similar to the previous case. \qed \medskip

\begin{corollary}
For all $1\leq m \leq n''$ we have
\beq \label{cKmain}
   2\cK(r_m) \geq
   D \biggl[ m + C_1' \ln m + C_2' m
   \Bigl( \frac mn \Bigr )^{\frac{2\beta}{\beta-2}}
   + C_3'\biggr]^{-2}
\eeq
where $D=\frac{\beta(\beta-1)}{(\beta-2)^2}$ and
$C_1',C_2',C_3'>0$ are some constants.
\end{corollary}

\proof Equation (\ref{cKrm}) implies
$$
   \cK(r_m) = \beta(\beta-1)x_m^{\beta-2}
   +\cO\bigl(x_m^{3\beta-4}\bigr).
$$
To estimate the main term we use (\ref{xmain}), and the remainder
term is $\cO\bigl(m^{-\frac{6\beta-8}{\beta-2}}\bigr)$ by
(\ref{xwbeta}), so it can be incorporated into the right hand side
of (\ref{cKmain}) by choosing sufficiently large constants
$C_1',C_2',C_3'>0$. \qed

\begin{lemma} \label{LmB}
For all $1\leq m < n''$ we have
\beq \label{cBmain}
   \cB(X_{m-1}) \geq
   A\biggl[ m + C_4 \ln m + C_5 m
   \Bigl( \frac mn \Bigr )^{\frac{2\beta}{\beta-2}}
   + C_6\biggr]^{-1}
\eeq
where $A>0$ satisfies $2A^2 - A = D$, hence
$A=\frac{\beta-1}{\beta-2}$, and $C_4,C_5,C_6>0$ are large
constants.
\end{lemma}

\proof We use induction on $m$. For $m=1$ the validity of
(\ref{cBmain}) is guaranteed by choosing $C_6$ large enough.
Assume that (\ref{cBmain}) is valid for some $m <n''-1$. Due to
(\ref{cBXm}) and (\ref{cKmain}) it is enough to verify
\begin{align*}
   \frac{D}{\Bigl[ m + C_1' \ln m + C_2' m
   \bigl( \frac mn \bigr )^{\frac{2\beta}{\beta-2}}
   + C_3'\Bigr]^{2}}+
   \frac{A}{ A\tau_m + m + C_4 \ln m + C_5 m
   \bigl( \frac mn \bigr )^{\frac{2\beta}{\beta-2}}
   + C_6} \\   >
   \frac{A}{ m+1 + C_4 \ln (m+1) + C_5 (m+1)
   \bigl( \frac{m+1}{n} \bigr )^{\frac{2\beta}{\beta-2}}
   + C_6}
\end{align*}
provided $C_4,C_5,C_6>0$ are large enough. Here
$$
   \tau_m = \tau(X_{m}) = 2+\cO(w_m)
   = 2+\cO\bigl(m^{\frac{\beta}{\beta-2}}\bigr).
$$
It is easy to see that
\begin{align*}
   \frac{A}{ m+1 + C_4 \ln (m+1) + C_5 (m+1)
   \bigl( \frac{m+1}{n} \bigr )^{\frac{2\beta}{\beta-2}}
   + C_6} \\
   -\frac{A}{ A\tau_m + m + C_4 \ln m + C_5 m
   \bigl( \frac mn \bigr )^{\frac{2\beta}{\beta-2}}
   + C_6} < \frac{2A^2-A}{\Theta}
\end{align*}
where $\Theta$ denotes the product of the two denominators. Thus
it is enough to verify
$$
   \frac{D}{\Bigl[ m + C_1' \ln m + C_2' m
   \bigl( \frac mn \bigr )^{\frac{2\beta}{\beta-2}}
   + C_3'\Bigr]^{2}} > \frac{2A^2-A}{\Theta}.
$$
We recall that $2A^2 - A = D$. Thus it is enough to verify
\beq \label{Theta}
   \Theta > \biggl[ m + C_1' \ln m + C_2' m
   \Bigl( \frac mn \Bigr )^{\frac{2\beta}{\beta-2}}
   + C_3'\biggr]^{2}.
\eeq
The leading term $m^2$ appears on both sides and cancels out.
Keeping only the largest non-cancelling terms on both sides of
(\ref{Theta}) we obtain
$$
  2\,C_4m\ln m + 2\,C_5m^2\Bigl( \frac mn
  \Bigr )^{\frac{2\beta}{\beta-2}}
  >
  2\,C_1'm\ln m + 2\,C_2'm^2\Bigl( \frac mn
  \Bigr )^{\frac{2\beta}{\beta-2}},
$$
which can be ensured by choosing $C_4$ and $C_5$ large enough.
This implies (\ref{Theta}) and then Lemma~\ref{LmB}. \qed

\begin{corollary}
\beq \label{cBmain2}
    \cB(X_{m-1})\geq
   \frac Am + \frac{C_4'\ln m}{m^2} + \frac{C_5'm}{n^2}
   + \frac{C_6'}{m^2},
\eeq
where $C_4',C_5',C_6'>0$ are large constants.
\end{corollary}

\proof This follows from (\ref{cBmain}) by Taylor expansion and
because $\tfrac{2\beta}{\beta-2}>2$. \qed \medskip

Now we are ready to estimate the expansion factor $\Lambda_n(X)$
given by (\ref{Lambdan}).

\begin{lemma}
We have
\beq \label{Lambdan''}
     \prod_{m=0}^{n''-1} \bigl(1 + \tau(X_m) \cB(X_m)\bigr)
     \geq C n^{\frac{2\beta-2}{\beta-2}}
\eeq
where $C>0$ is a constant.
\end{lemma}

\proof Note that $\tau(X_m) > 2$. Hence, due (\ref{cBmain2}), we
have
$$
     \ln \Biggl[\prod_{m=0}^{n''-1} \bigl(1 + \tau(X_m)
     \cB(X_m)\bigr)\Biggr]
     >\sum_{m=1}^{n''} \biggl[\frac{2A}{m} +
     \frac{2C_4'\,\ln m}{m^2} + \frac{2C_5'm}{n^2}
   + \frac{C_7}{m^2}\biggr]
$$
with some large constant $C_7>0$. Therefore,
$$
     \ln \Biggl[\prod_{m=0}^{n''-1} \bigl(1 + \tau(X_m)
     \cB(X_m)\bigr)\Biggr]
     >2A\ln n'' + \text{const} > 2A\ln n +\text{const},
$$
where the last inequality follows from (\ref{nnn}). Lastly, note
that $2A =\frac{2\beta-2}{\beta-2}$, which completes the proof of
the lemma. \qed
\medskip

The bound (\ref{Lambdan''}) implies
\beq \label{Lambdan1}
     \Lambda_n^{(1)}(X)\colon=
     \prod_{m=0}^{n'-1} \bigl(1 + \tau(X_m) \cB(X_m)\bigr)
     \geq C n^{\frac{2\beta-2}{\beta-2}}
\eeq

\begin{lemma}
\beq \label{Lambdan2}
     \Lambda_n^{(2)}(X)\colon=
     \prod_{m=n'}^{n-1} \bigl(1 + \tau(X_m) \cB(X_m)\bigr)
     \geq C n^{\frac{\beta}{\beta-2}}
\eeq
where $C>0$ is a constant.
\end{lemma}

\proof This can be obtained by a detailed analysis of the dynamics
on the interval $(n',n)$ similar to the one done for the interval
$(0,n')$, but we will use a shortcut: the time-reversibility of
the billiard dynamics will allow us to derive (\ref{Lambdan2})
directly from (\ref{Lambdan1}).

Let $V^u$ and $V^s$ be two unit vectors tangent to the unstable
and stable manifolds, respectively, at the point $X$. Since the
angle between $V^u$ and $V^s$ is bounded away from zero, the area
of the parallelogram $\Pi$ spanned by $V^u$ and $V^s$ is of order
one (uniformly in $n$).

\begin{figure}[h]
\centering \psfrag{P}{$\Pi$} \psfrag{F}[][]{$D_X\cF^{n'}$}
\psfrag{u}{$V^u$} \psfrag{s}{$V^s$} \psfrag{X}{$X$}
\psfrag{Q}{$\Pi_{n'}$} \psfrag{O}{$\cO(\tfrac 1n)$}
\includegraphics[width=5in,height=1in]{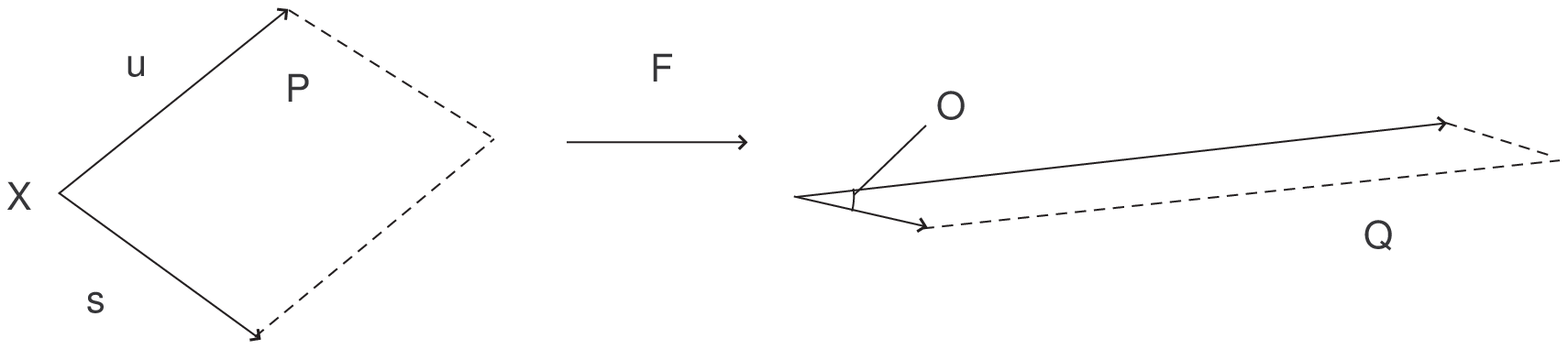}
\renewcommand{\figurename}{Fig.}
\caption{}\label{FigFn}
\end{figure}

Consider the parallelogram $\Pi_{n'} = D_X\cF^{n'} (\Pi)$ spanned
by the vectors $V_{n'}^u = D_X\cF^{n'} (V^u)$ and $V_{n'}^s =
D_X\cF^{n'} (V^s)$. Since the map $\cF^{n'}$ preserves the measure
$d\mu = \cos\varphi\, dr\, d\varphi$, we have
$$
   \cos\varphi_{n'}\,\, \text{Area}(\Pi_{n'}) =
   \cos\varphi\,\, \text{Area}(\Pi).
$$
Note that $\cos\varphi \approx 1$ and $\cos\varphi_{n'} \approx
1$, hence
$$
   \text{Area}(\Pi_{n'}) \sim \text{Area}(\Pi) \sim 1.
$$
On the other hand,
$$
   \text{Area}(\Pi_{n'}) = |V^u_{n'}|\,|V^s_{n'}|\,\sin\gamma_{n'}
$$
where $ |V^u_{n'}|$ and $|V^s_{n'}|$ denote the lengths of these
vectors in the Euclidean norm (\ref{Enorm}) and $\gamma_{n'}$
denotes the angle between them.

Next we estimate $\gamma_{n'}$. It easily follows from
(\ref{cBXm}) that
$$
   \cB(X_{n'})\geq \Biggl[\sum_{m=0}^{n'-1} \tau(X_m) +
   1/\cB(X_0)\Biggr]^{-1} \sim \frac{1}{n'} \sim \frac 1n.
$$
Now (\ref{cBY}) implies that the slope of the vector $V^u_{n'}$ is
$$
   \frac{d\varphi}{dr} = \cos\varphi_{n'}\,\,\cB(X_{n'}) -
   \cK(r_{n'}).
$$
We note that $\cos\varphi_{n'} \approx 1$ and $\cK(r_{n'}) \sim
n^{-\beta}$ due to (\ref{cKrm}), because $x_{n'} <3 w_{n'} \sim
n^{\frac{\beta}{2-\beta}}$, cf.\ (\ref{n'n''asym}). Therefore,
$$
    \frac{d\varphi}{dr} > \frac{C}{n}
$$
for some constant $C>0$. Hence the vector $V^u_{n'}$ makes an
angle $\geq C/n$ with the horizontal $r$-axis. By the time
reversibility, the vector $V^s_{n'}$ makes an angle $\leq -C/n$
with the horizontal $r$-axis, see Fig.~\ref{FigFn}, hence
$\sin\gamma_{n'} > c/n$ for some constant $c>0$, and we obtain
$$
   |V^u_{n'}|\,|V^s_{n'}| < cn
$$
for some constant $c>0$.

Next, the Euclidean norm $|V|$ defined by (\ref{Enorm}) is
uniformly equivalent to the p-norm (\ref{pnorm}) for both stable
and unstable vectors in our considerations. Indeed, $\cos\varphi
\approx 1$ and $|d\varphi| \leq C\, |dr|$ for some constant $C>0$,
as it easily follows from (\ref{cBY}). Therefore, we obtain
$$
   |V^u_{n'}|_p\,|V^s_{n'}|_p < cn
$$
for some constant $c>0$. Obviously,
$$
  |V^u_{n'}|_p = \Lambda_{n'}^{(1)}(X) \,  |V^u|_p
  \sim \Lambda_{n'}^{(1)}(X).
$$
Now it is time for a little trick. By the time reversibility of
the billiard dynamics, the contraction of stable vectors during
the time interval $(0,n')$ is the same as the expansion of the
corresponding unstable vectors during the time interval $(n',n)$,
hence
$$
  |V^s_{n'}|_p \sim \bigl[ \Lambda_{n'}^{(2)}(X)\bigr]^{-1} |V^s|_p
  \sim  \bigl[ \Lambda_{n'}^{(2)}(X)\bigr]^{-1}.
$$
Therefore,
\beq \label{Lambda12}
   \Lambda_{n'}^{(2)}(X) > c \Lambda_{n'}^{(1)}(X)/n
\eeq
for some constant $c>0$. Now (\ref{Lambda12}) and (\ref{Lambdan1})
imply (\ref{Lambdan2}). \qed \medskip

Combining (\ref{Lambdan1}) and (\ref{Lambdan2}) gives
$$
  \Lambda_n(X) =  \Lambda_n^{(1)}(X)\, \Lambda_n^{(2)}(X)
  \geq C n^{\frac{3\beta-2}{\beta-2}}.
$$
This proves Proposition~\ref{PrAux} for $W\subset C_n''$ because,
in its notation, we have
$$
   b=a+2=\frac{3\beta-2}{\beta-2}.
$$

We now consider the remaining case $W \subset C_n'$, which
correspond to trajectories that start near the point $q_1$, enter
the window $|x|<\varepsilon$, but turn around before reaching the
central line $x=0$ and come back into the vicinity of $q_1$ or
$q_2$ (as shown by the solid line on Fig.~\ref{FigWindow}).

In that case $n'$ can be defined as the turning point, i.e.\ by
$x_{n'}<x_{n'-1}$ and $x_{n'}<x_{n'+1}$. Observe that if
$X'=(r',\varphi')\in C_n'$, then there exists another point
$X=(r,\varphi)\in C_n''$ with $r=r'$ and $\varphi< \varphi'$,
whose trajectory goes through the window, as it is clear from
Fig.~\ref{FigSing}. Since $\varphi'< \varphi$, it follows that the
$x$-coordinate $x_m$ of the point $\cF^m(X)$ will be always
smaller than the $x$-coordinate $x_m'$ of the point $\cF^m(X')$,
for all $1\leq m\leq n$. This observation and the bound
(\ref{xmain}) that we have proved for $x_m$ implies that the same
bound holds for $x_m'$ and for all $1\leq m\leq n''$. The rest of
the proof of Proposition~\ref{PrAux} for $X'\in C_n'$ is identical
to that of the case $X\in C_n''$.

Proposition~\ref{PrAux} is now proven. \qed

\end{document}